\documentclass[12pt,a4paper]{article}
\usepackage{ifthen} 
\newboolean{pdflatex}
\setboolean{pdflatex}{true}  

\newboolean{articletitles}
\setboolean{articletitles}{true}

\newboolean{uprightparticles}
\setboolean{uprightparticles}{false} 

\def\paperauthors{LHCb Collaboration} 
\def\paperasciititle{Evidence of a Jpsi Lambda  resonance  //  and observation of excited Xi- states //  in the Xib -> Jpsi Lambda K- decay}  
\def\paperkeywords{{High Energy Physics}, {LHCb}} 
\def\papercopyright{\the\year\ CERN for the benefit of the LHCb Collaboration}
\def\paperlicence{CC BY 4.0 licence}
\def\paperlicenceurl{https://creativecommons.org/licenses/by/4.0/}

\usepackage[top=1in, bottom=1.25in, left=1in, right=1in]{geometry}

\usepackage{microtype}
\usepackage{lineno}  
\usepackage{xspace} 
\usepackage{caption} 
\renewcommand{\captionfont}{\small}
\renewcommand{\captionlabelfont}{\small}

\usepackage{graphicx}  
\usepackage{color}
\usepackage{colortbl}
\graphicspath{{./figs/}} 

\usepackage{amsmath} 
\usepackage{amssymb}
\usepackage{amsfonts}
\usepackage{upgreek} 
\usepackage{floatrow}
\usepackage[label font=bf]{subfig}
\newcommand*\patchAmsMathEnvironmentForLineno[1]{%
\expandafter\let\csname old#1\expandafter\endcsname\csname #1\endcsname
\expandafter\let\csname oldend#1\expandafter\endcsname\csname
end#1\endcsname
 \renewenvironment{#1}%
   {\linenomath\csname old#1\endcsname}%
   {\csname oldend#1\endcsname\endlinenomath}%
}
\newcommand*\patchBothAmsMathEnvironmentsForLineno[1]{%
  \patchAmsMathEnvironmentForLineno{#1}%
  \patchAmsMathEnvironmentForLineno{#1*}%
}
\AtBeginDocument{%
\patchBothAmsMathEnvironmentsForLineno{equation}%
\patchBothAmsMathEnvironmentsForLineno{align}%
\patchBothAmsMathEnvironmentsForLineno{flalign}%
\patchBothAmsMathEnvironmentsForLineno{alignat}%
\patchBothAmsMathEnvironmentsForLineno{gather}%
\patchBothAmsMathEnvironmentsForLineno{multline}%
\patchBothAmsMathEnvironmentsForLineno{eqnarray}%
}

\usepackage{hyperxmp}

\usepackage[pdftex,
            pdfauthor={\paperauthors},
            pdftitle={\paperasciititle},
            pdfkeywords={\paperkeywords},
            pdfcopyright={Copyright (C) \papercopyright},
            pdflicenseurl={\paperlicenceurl}]{hyperref}

\usepackage[colorinlistoftodos,textsize=scriptsize]{todonotes}

\usepackage[bottom,flushmargin,hang,multiple]{footmisc}

\usepackage[all]{hypcap} 

\usepackage{xspace} 
\usepackage{upgreek}

\def\lhcb   {\mbox{LHCb}\xspace}

\def\MagUp {\mbox{\em Mag\kern -0.05em Up}\xspace}

\ifthenelse{\boolean{uprightparticles}}%
{

 \def\Pmu         {\ensuremath{\upmu}\xspace}

 \def\Ppi         {\ensuremath{\uppi}\xspace}

 \def\Ppsi        {\ensuremath{\uppsi}\xspace}

 \def\PDelta      {\ensuremath{\Delta}\xspace}                 
 \def\PXi         {\ensuremath{\Xi}\xspace}                 
 \def\PLambda     {\ensuremath{\Lambda}\xspace}                 
 \def\PSigma      {\ensuremath{\Sigma}\xspace}                 
 \def\POmega      {\ensuremath{\Omega}\xspace}                 
 \def\PUpsilon    {\ensuremath{\Upsilon}\xspace}

 \def\PB      {\ensuremath{\mathrm{B}}\xspace}                 
                  
 \def\PD      {\ensuremath{\mathrm{D}}\xspace}

 \def\PJ      {\ensuremath{\mathrm{J}}\xspace}                 
 \def\PK      {\ensuremath{\mathrm{K}}\xspace}

 \def\Pb      {\ensuremath{\mathrm{b}}\xspace}                 
 \def\Pc      {\ensuremath{\mathrm{c}}\xspace}

 \def\Pi      {\ensuremath{\mathrm{i}}\xspace}

 \def\Pp      {\ensuremath{\mathrm{p}}\xspace}

 \def\Ps      {\ensuremath{\mathrm{s}}\xspace}

 \def\thebaroffset{0.0em}
}
{

 \def\Pmu         {\ensuremath{\mu}\xspace}

 \def\Ppi         {\ensuremath{\pi}\xspace}

 \def\Ppsi        {\ensuremath{\psi}\xspace}                 
                  
 \mathchardef\PDelta="7101
 \mathchardef\PXi="7104
 \mathchardef\PLambda="7103
 \mathchardef\PSigma="7106
 \mathchardef\POmega="710A
 \mathchardef\PUpsilon="7107
                  
 \def\PB      {\ensuremath{B}\xspace}                 
                  
 \def\PD      {\ensuremath{D}\xspace}

 \def\PJ      {\ensuremath{J}\xspace}                 
 \def\PK      {\ensuremath{K}\xspace}

 \def\Pb      {\ensuremath{b}\xspace}                 
 \def\Pc      {\ensuremath{c}\xspace}

 \def\Pi      {\ensuremath{i}\xspace}

 \def\Pp      {\ensuremath{p}\xspace}

 \def\Ps      {\ensuremath{s}\xspace}

 \def\thebaroffset{0.18em}
}
\newcommand{\offsetoverline}[2][\thebaroffset]{\kern #1\overline{\kern -#1 #2}}%

\makeatletter
\ifcase \@ptsize \relax
  \newcommand{\miniscule}{\@setfontsize\miniscule{4}{5}}
\or
  \newcommand{\miniscule}{\@setfontsize\miniscule{5}{6}}
\or
  \newcommand{\miniscule}{\@setfontsize\miniscule{5}{6}}
\fi
\makeatother

\DeclareRobustCommand{\optbar}[1]{\shortstack{{\miniscule (\rule[.5ex]{1.25em}{.18mm})}
  \\ [-.7ex] $#1$}}

\def\mup        {{\ensuremath{\Pmu^+}}\xspace}
\def\mun        {{\ensuremath{\Pmu^-}}\xspace} 

\def\squark    {{\ensuremath{\Ps}}\xspace}

\def\cquark    {{\ensuremath{\Pc}}\xspace}

\def\bquark    {{\ensuremath{\Pb}}\xspace}

\def\pion   {{\ensuremath{\Ppi}}\xspace}

\def\pim    {{\ensuremath{\pion^-}}\xspace}

\def\kaon    {{\ensuremath{\PK}}\xspace}

\def\KorKbar {\kern \thebaroffset\optbar{\kern -\thebaroffset \PK}{}\xspace}

\def\Km      {{\ensuremath{\kaon^-}}\xspace}

\def\Dbar    {{\ensuremath{\offsetoverline{\PD}}}\xspace}
\def\D       {{\ensuremath{\PD}}\xspace}

\def\DorDbar {\kern \thebaroffset\optbar{\kern -\thebaroffset \PD}\xspace}

\def\Dp      {{\ensuremath{\D^+}}\xspace}
\def\Dm      {{\ensuremath{\D^-}}\xspace}

\def\DpDm    {\ensuremath{\Dp {\kern -0.16em \Dm}}\xspace}

\def\Dstarzb {{\ensuremath{\Dbar{}^{*0}}}\xspace}

\def\B       {{\ensuremath{\PB}}\xspace}

\def\BorBbar {\kern \thebaroffset\optbar{\kern -\thebaroffset \PB}\xspace}

\def\Bd      {{\ensuremath{\B^0}}\xspace}

\def\BdorBdbar {\kern \thebaroffset\optbar{\kern -\thebaroffset \Bd}\xspace}

\def\Bs      {{\ensuremath{\B^0_\squark}}\xspace}

\def\BsorBsbar {\kern \thebaroffset\optbar{\kern -\thebaroffset \Bs}\xspace}

\def\jpsi     {{\ensuremath{{\PJ\mskip -3mu/\mskip -2mu\Ppsi}}}\xspace}

\def\Y#1S{\ensuremath{\PUpsilon{(#1S)}}\xspace}

\def\proton      {{\ensuremath{\Pp}}\xspace}

\def\Lz          {{\ensuremath{\PLambda}}\xspace}

\def\LorLbar     {\kern \thebaroffset\optbar{\kern -\thebaroffset \PLambda}\xspace}

\def\Xires       {{\ensuremath{\PXi}}\xspace}

\def\Xim         {{\ensuremath{\Xires^-}}\xspace}

\def\Xicp        {{\ensuremath{\Xires^+_\cquark}}\xspace}

\def\Lb           {{\ensuremath{\Lz^0_\bquark}}\xspace}

\def\Xibm         {{\ensuremath{\Xires^-_\bquark}}\xspace}

\newcommand{\decay}[2]{\ensuremath{#1\!\to #2}\xspace} 

\def\to                 {\ensuremath{\rightarrow}\xspace}

\def\AT#1     {\ensuremath{A_{\mathrm{T}}^{#1}}\xspace}           

\def\C#1      {\ensuremath{\mathcal{C}_{#1}}\xspace}                       
\def\Cp#1     {\ensuremath{\mathcal{C}_{#1}^{'}}\xspace}                    
\def\Ceff#1   {\ensuremath{\mathcal{C}_{#1}^{\mathrm{(eff)}}}\xspace}        
\def\Cpeff#1  {\ensuremath{\mathcal{C}_{#1}^{'\mathrm{(eff)}}}\xspace}       
\def\Ope#1    {\ensuremath{\mathcal{O}_{#1}}\xspace}                       
\def\Opep#1   {\ensuremath{\mathcal{O}_{#1}^{'}}\xspace}                    

\newcommand{\aunit}[1]{\ensuremath{\text{\,#1}}}       

\newcommand{\tev}{\aunit{Te\kern -0.1em V}\xspace}
\newcommand{\gev}{\aunit{Ge\kern -0.1em V}\xspace}
\newcommand{\mev}{\aunit{Me\kern -0.1em V}\xspace}
\newcommand{\kev}{\aunit{ke\kern -0.1em V}\xspace}
\newcommand{\ev}{\aunit{e\kern -0.1em V}\xspace}
 
\newcommand{\mevc}{\ensuremath{\aunit{Me\kern -0.1em V\!/}c}\xspace}
\newcommand{\gevc}{\ensuremath{\aunit{Ge\kern -0.1em V\!/}c}\xspace}
\newcommand{\mevcc}{\ensuremath{\aunit{Me\kern -0.1em V\!/}c^2}\xspace}
\newcommand{\gevcc}{\ensuremath{\aunit{Ge\kern -0.1em V\!/}c^2}\xspace}

\def\m    {\aunit{m}\xspace}

\def\fb   {\ensuremath{\aunit{fb}}\xspace}
\def\invfb   {\ensuremath{\fb^{-1}}\xspace}

\newcommand{\stat}{\aunit{(stat)}\xspace}

\newcommand{\chisq}{\ensuremath{\chi^2}\xspace}

\newcommand{\chisqip}{\ensuremath{\chi^2_{\text{IP}}}\xspace}

\def\gsim{{~\raise.15em\hbox{$>$}\kern-.85em
          \lower.35em\hbox{$\sim$}~}\xspace}
\def\lsim{{~\raise.15em\hbox{$<$}\kern-.85em
          \lower.35em\hbox{$\sim$}~}\xspace}

\def\pt         {\ensuremath{p_{\mathrm{T}}}\xspace}

\def\evtgen     {\mbox{\textsc{EvtGen}}\xspace}

\def\geant      {\mbox{\textsc{Geant4}}\xspace}

\def\photos     {\mbox{\textsc{Photos}}\xspace}

\def\pythia     {\mbox{\textsc{Pythia}}\xspace}

\def\tell1  {TELL1\xspace}
\def\ukl1   {UKL1\xspace}

\usepackage{cite} 
\usepackage{mciteplus}

\newcommand{\mygevc}{\ensuremath{{\mathrm{Ge\kern -0.1em V\!/}c}}\xspace}
\newcommand{\xx}{\ensuremath{\kern 0.5em }}

\newcommand{\LbJpsipK}{\ensuremath{\Lb\to\jpsi\proton\Km}\xspace}

\newcommand{\LbJpsiL}{\ensuremath{\Lb\to\jpsi\Lz}\xspace}

\newcommand{\Jpsip}{\ensuremath{\jpsi\proton}\xspace}
\newcommand{\JpsiL}{\ensuremath{\jpsi\Lz}\xspace}
\newcommand{\JpsiLK}{\ensuremath{\jpsi\Lz\Km}\xspace}
\newcommand{\LK}{\ensuremath{\Lz\Km}\xspace}

\newcommand{\XbJpsiLK}{\ensuremath{\Xibm\to\jpsi\Lz\Km}\xspace}
\newcommand{\XbJpsiSK}{\ensuremath{\Xibm\to\jpsi\PSigma^{0}(\to\Lz\gamma)\Km}\xspace}

\newcommand{\Xistm}{\ensuremath{\PXi^{*-}}\xspace}

\def\Like{\mathcal{L}}

\def\pcs{\ensuremath{P_{cs}}\xspace}
\def\Pcs{\ensuremath{P_{cs}^0}\xspace}
\def\TPcs{\ensuremath{P_{cs}(4459)^0}\xspace}
\def\mPc{\ensuremath{m_{\jpsi\Lz}}\xspace}
\def\mLK{\ensuremath{m_{\Lz\Km}}\xspace}

\def\Xione{\ensuremath{\PXi(1690)^-}\xspace}
\def\Xitwo{\ensuremath{\PXi(1820)^-}\xspace}
\def\Xithree{\ensuremath{\PXi(1950)^-}\xspace}
\def\Xifour{\ensuremath{\PXi(2030)^-}\xspace}

\usepackage{multirow} 
\usepackage{booktabs} 
\usepackage{rotating}

\usepackage{longtable}

\begin{document}

\renewcommand{\thefootnote}{\fnsymbol{footnote}}
\setcounter{footnote}{1}

\begin{titlepage}
\pagenumbering{roman}

\vspace*{-1.5cm}
\centerline{\large EUROPEAN ORGANIZATION FOR NUCLEAR RESEARCH (CERN)}
\vspace*{1.5cm}
\noindent
\begin{tabular*}{\linewidth}{lc@{\extracolsep{\fill}}r@{\extracolsep{0pt}}}
\ifthenelse{\boolean{pdflatex}}
{\vspace*{-1.5cm}\mbox{\!\!\!\includegraphics[width=.14\textwidth]{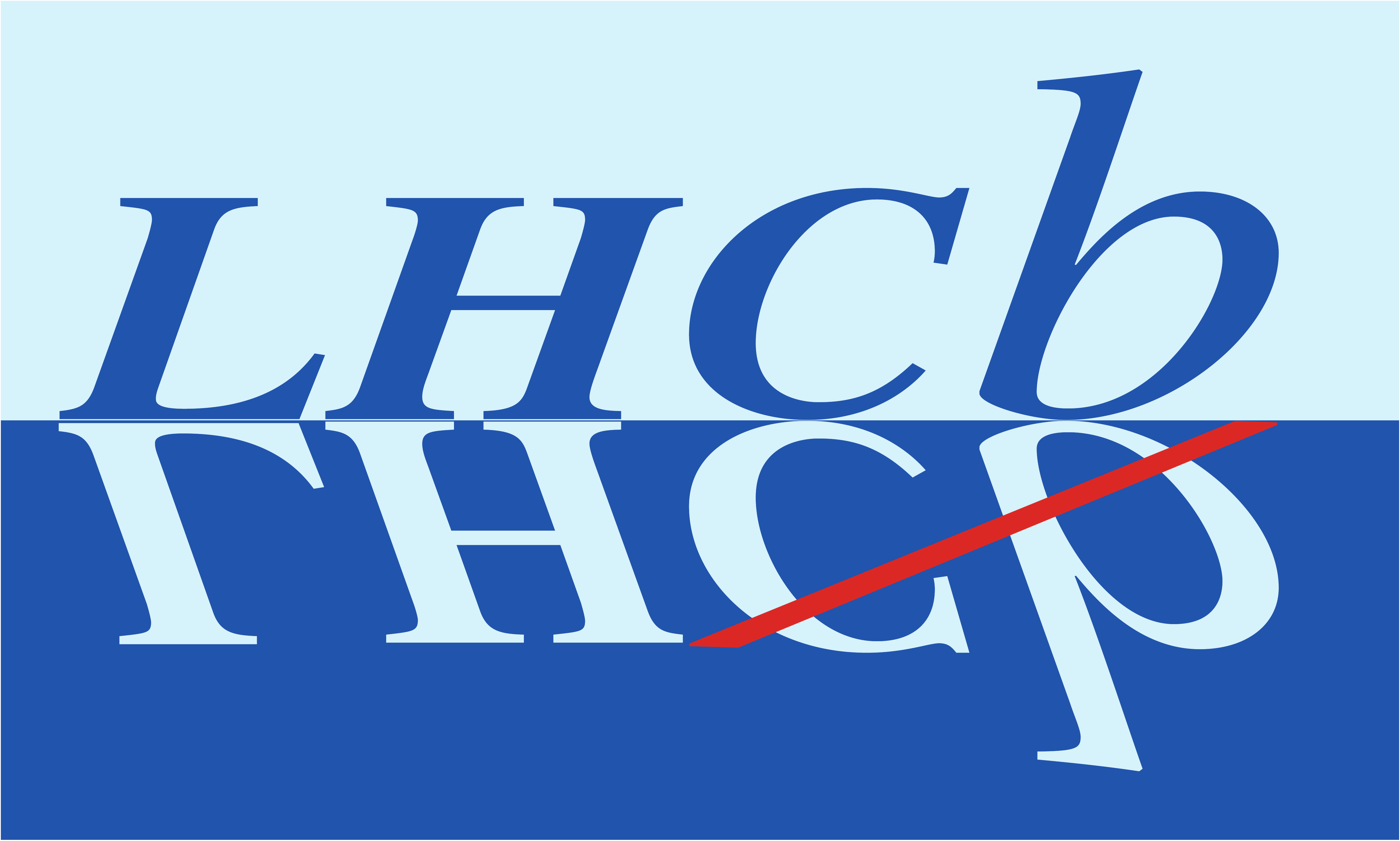}} & &}
{\vspace*{-1.2cm}\mbox{\!\!\!\includegraphics[width=.12\textwidth]{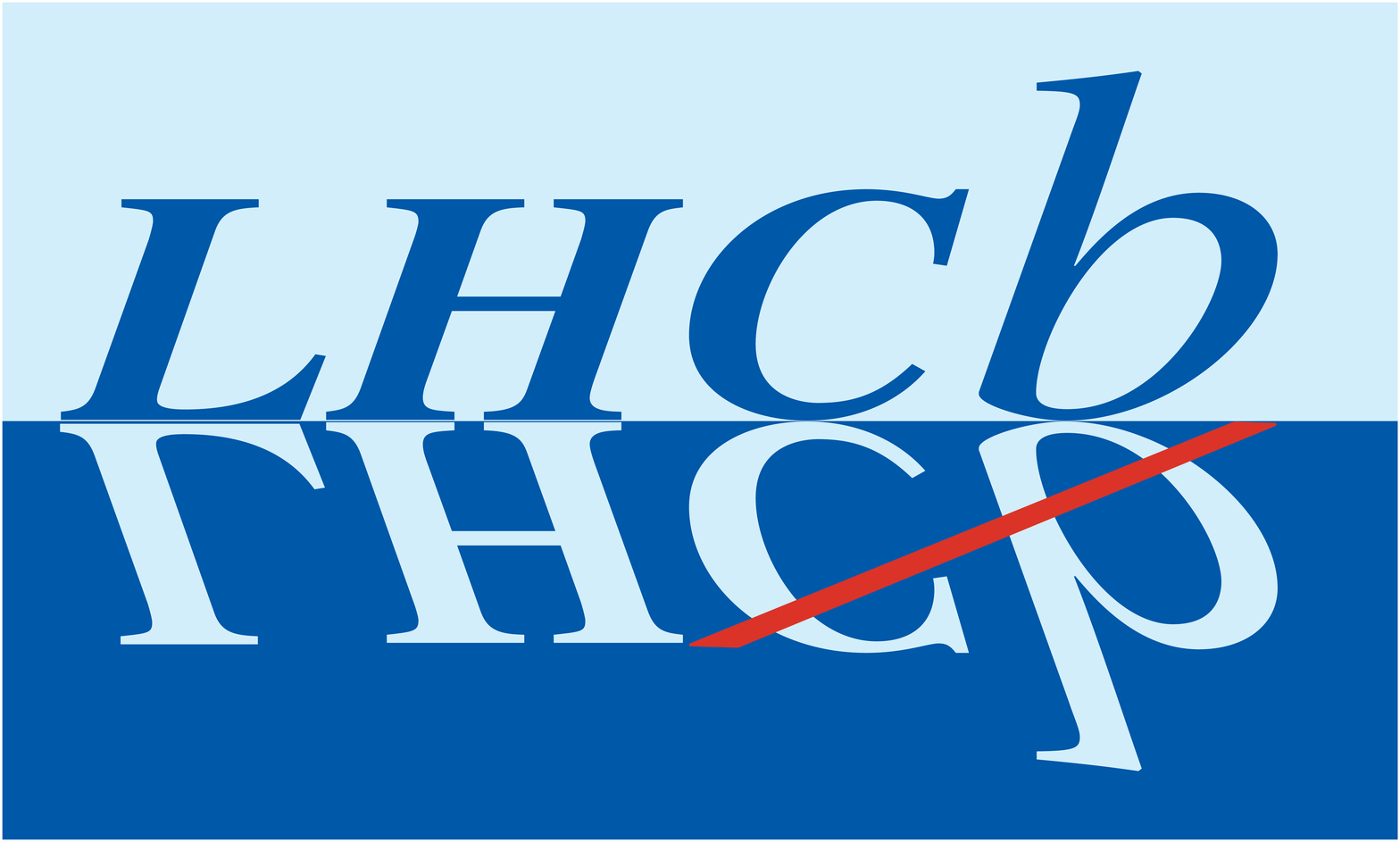}} & &}
\\
 & & CERN-EP-2020-233 \\  
 & & LHCb-PAPER-2020-039 \\   
 & & 3 July 2021 \\
\end{tabular*}

\vspace*{3.0cm}

{\normalfont\bfseries\boldmath\huge
\begin{center}
  {Evidence of a $\jpsi\Lz$ structure \\and observation of excited $\PXi^-$ states in the $\Xibm\to\jpsi\Lz\Km$ decay} 
\end{center}
}

\vspace*{2.0cm}

\begin{center}
\paperauthors\footnote{Authors are listed at the end of this paper.}
\end{center}

\vspace{\fill}

\begin{abstract}
  \noindent

  First evidence of a structure in the $\jpsi\Lz$ invariant mass distribution is obtained from an amplitude analysis of \mbox{$\XbJpsiLK$} decays. The observed structure is consistent with being due to a charmonium pentaquark with strangeness with a significance of 3.1$\sigma$ including systematic uncertainties and look-elsewhere effect.
Its mass and width are determined to be $4458.8\pm2.9\,^{+4.7}_{-1.1}\mev$ and $17.3\pm6.5\,^{+8.0}_{-5.7}\mev$, respectively, where the quoted uncertainties are statistical and systematic. The structure is also consistent with being due to two resonances. In addition, the narrow excited $\PXi^-$ states, $\Xione$ and $\Xitwo$, are seen for the first time in a $\Xibm$ decay, and their masses and widths are measured with improved precision. The analysis is performed using $pp$ collision data corresponding to a total integrated luminosity of 9\invfb, collected with the LHCb experiment at centre-of-mass energies of $7$, $8$ and $13$\tev.

\end{abstract}

\vspace*{1.0cm}

Keyswords: QCD; exotics; pentaquark; spectroscopy; quarkonium; particle and resonance production

\begin{center}
  Published in
  Science Bulletin 66 (2021) 1278-1287
\end{center}

\vspace{\fill}

{\footnotesize 
\centerline{\copyright~\papercopyright. \href{\paperlicenceurl}{\paperlicence}.}}
\vspace*{2mm}

\end{titlepage}

\newpage
\setcounter{page}{2}
\mbox{~}

\renewcommand{\thefootnote}{\arabic{footnote}}
\setcounter{footnote}{0}

\cleardoublepage

\pagestyle{plain} 
\setcounter{page}{1}
\pagenumbering{arabic}

\section{Introduction}

The existence of pentaquark states, comprising four quarks and an antiquark, has been anticipated since the establishment of the quark model~\cite{GellMann:1964nj,Zweig:352337}. The first observation of pentaquark states has been reported by the LHCb experiment~\cite{LHCb-PAPER-2015-029}. In the analysis of $\Lb\to\jpsi(\to\mup\mu^-)pK^-$ decays, a narrow structure in the \Jpsip mass spectrum indicated a possible pentaquark contribution~\cite{LHCb-PAPER-2015-029,LHCb-PAPER-2016-009}. 
(Charge conjugation is implied and natural units with $\hbar=c=1$ are used throughout this article.) An amplitude analysis showed that the data could be best described by the presence of two pentaquark states, the $P_c(4380)^+$ and $P_c(4450)^+$. With the inclusion of additional data and an improved selection strategy, it was found that the $P_c(4450)^+$ could be resolved into two narrow states, the $P_c(4440)^+$ and $P_c(4457)^+$. In addition, a new narrow state, the  $P_c(4312)^+$ was  discovered~\cite{LHCb-PAPER-2019-014}. The $[uudc\bar{c}]$ valence quark content is attributed to these pentaquark states. Their strange counterparts, denoted $P_{cs}^0$, with $[udsc\bar{c}]$ valence quark content, are predicted in Refs.~\cite{Wu:2010jy,Chen:2016ryt,Santopinto:2016pkp,Shen:2019evi,Xiao:2019gjd,Wang:2019nvm} and it has been suggested to search for them in \XbJpsiLK decays~\cite{Chen:2015sxa,Santopinto:2016pkp}.

The \XbJpsiLK decay also provides the opportunity to study  excited \Xim  resonances (denoted collectively as \Xistm) in a mass range of  $[1.61, 2.70]\gev$, where only a small number of \Xistm states have been observed, 
with a typical uncertainty of more than $5 \mev$ on their masses and widths~\cite{PDG2020}. Five $\Xistm$ states have been established experimentally, the $\PXi(1530)^-$, $\PXi(1690)^-$, $\PXi(1820)^-$,  $\PXi(1950)^-$ and $\PXi(2030)^-$.
Recently, several results on the \Xione and \Xitwo states have been reported by the BESIII Collaboration~\cite{Ablikim:2015apm,Ablikim:2019kkp}. For the neutral partners, the Belle Collaboration observed a  $\varXi(1620)^0$ resonance and found evidence for a $\varXi(1690)^0$ state in $\Xicp\to \PXi^- \pi^+\pi^+$ decays~\cite{Sumihama:2018moz}. More studies of excited $\varXi$ states will improve our understanding of the $\varXi$ spectrum and of the structure of baryon resonances.

In this article, an amplitude analysis of the \XbJpsiLK decay is performed using proton-proton ($pp$) collision data  collected at centre-of-mass energies of \mbox{$\sqrt{s}=7$ and 8\tev}, corresponding to an integrated luminosity of 3\invfb (Run~1) and at $\sqrt{s} = 13\tev$, corresponding to 6\invfb (Run~2). The LHCb Collaboration first observed the decay \mbox{\XbJpsiLK} using Run 1 data and measured the production rate of \Xibm with \XbJpsiLK decays relative to that of $\decay{\Lb}{\jpsi\Lz}$ decays~\cite{LHCb-PAPER-2016-053}. 

\section{Detector and data set}
The LHCb detector~\cite{LHCb-DP-2008-001,LHCb-DP-2014-002} is a
single-arm forward spectrometer covering the pseudorapidity range $2 < \eta < 5$, designed for
the study of particles containing \bquark\ or \cquark\ quarks. The detector includes a
silicon-strip vertex detector surrounding the proton-proton interaction region, tracking
stations on either side of a dipole magnet, ring-imaging Cherenkov (RICH) detectors,
calorimeters and muon chambers. Simulation is required to evaluate the detector acceptance in the full phase space of \XbJpsiLK decays and the efficiency of signal selection. 
In the simulation, $pp$ collisions are generated using \pythia~\cite{Sjostrand:2007gs} with a specific \lhcb configuration~\cite{LHCb-PROC-2010-056}.
Decays of unstable particles are described by \evtgen~\cite{Lange:2001uf}, in which final-state radiation is generated using \photos~\cite{Golonka:2005pn,Davidson:2010ew}.
The interaction of the generated particles with the detector, and its response, are implemented using the \geant toolkit~\cite{Allison:2006ve,Agostinelli:2002hh} as described in Ref.~\cite{LHCb-PROC-2011-006}. The production kinematics of $\Xibm$ baryons in the simulation is corrected based on the two-dimensional distribution of the momentum component transverse to the beam direction (\pt) and rapidity ($y$) obtained using a sample of \LbJpsiL decays selected from data.

The $\Xibm\to\jpsi(\to\mup\mun)\Lz(\to \proton\pim) \Km$ signal candidates are first required to pass an online event selection performed by a trigger~\cite{LHCb-DP-2012-004}, consisting of a hardware stage that is based on information from the calorimeters and the muon system, followed by two software stages that perform a partial event reconstruction. At the hardware stage, events are required to have a muon with a high momentum component transverse to the beam direction (\pt) or a hadron, photon or electron with a high transverse energy. In the first software stage, the event is required to have either two well-identified oppositely charged muons with large invariant mass, or at least one muon with $\pt>1\gev$ and a large impact-parameter significance with respect to any primary \proton\proton collision  vertex (PV). In the second stage, events containing a $\mup\mun$ pair with invariant mass consistent with the known $\jpsi$ mass~\cite{PDG2020}, and with a vertex significantly displaced from any PV, are selected. Candidate  \decay{\Lz}{\proton\pim} decays are reconstructed in two different categories:
{\it long} involving \Lz baryons with a flight distance short enough for the proton and pion to have the decay vertex reconstructed in the vertex detector; 
and {\it downstream} containing \Lz baryons decaying later such that the track segments of the proton and pion cannot be formed in the vertex detector and are reconstructed only in the tracking stations. The candidates in the long category have better mass, momentum and decay vertex resolution than those in the downstream category. A $\Xibm$ candidate is then reconstructed by combining the $\jpsi$, $\Lz$ and a well identified $\Km$ candidate, which are required to form a good-quality vertex. 

\section{Candidate selection}

\begin{figure}[t]
\centering
\sidesubfloat[]{
\includegraphics[width=0.45\textwidth]{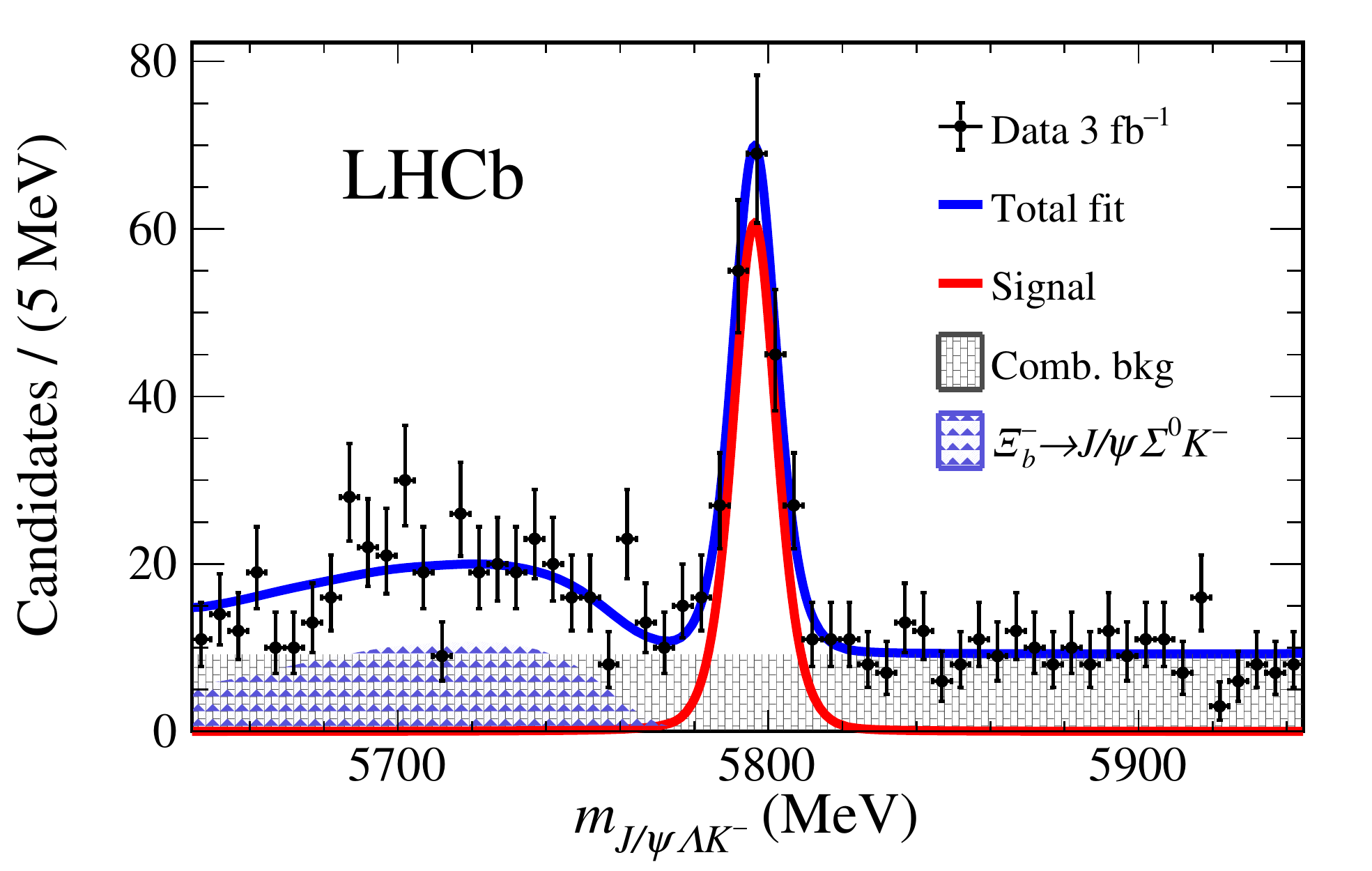}}%
\sidesubfloat[]{
\includegraphics[width=0.45\textwidth]{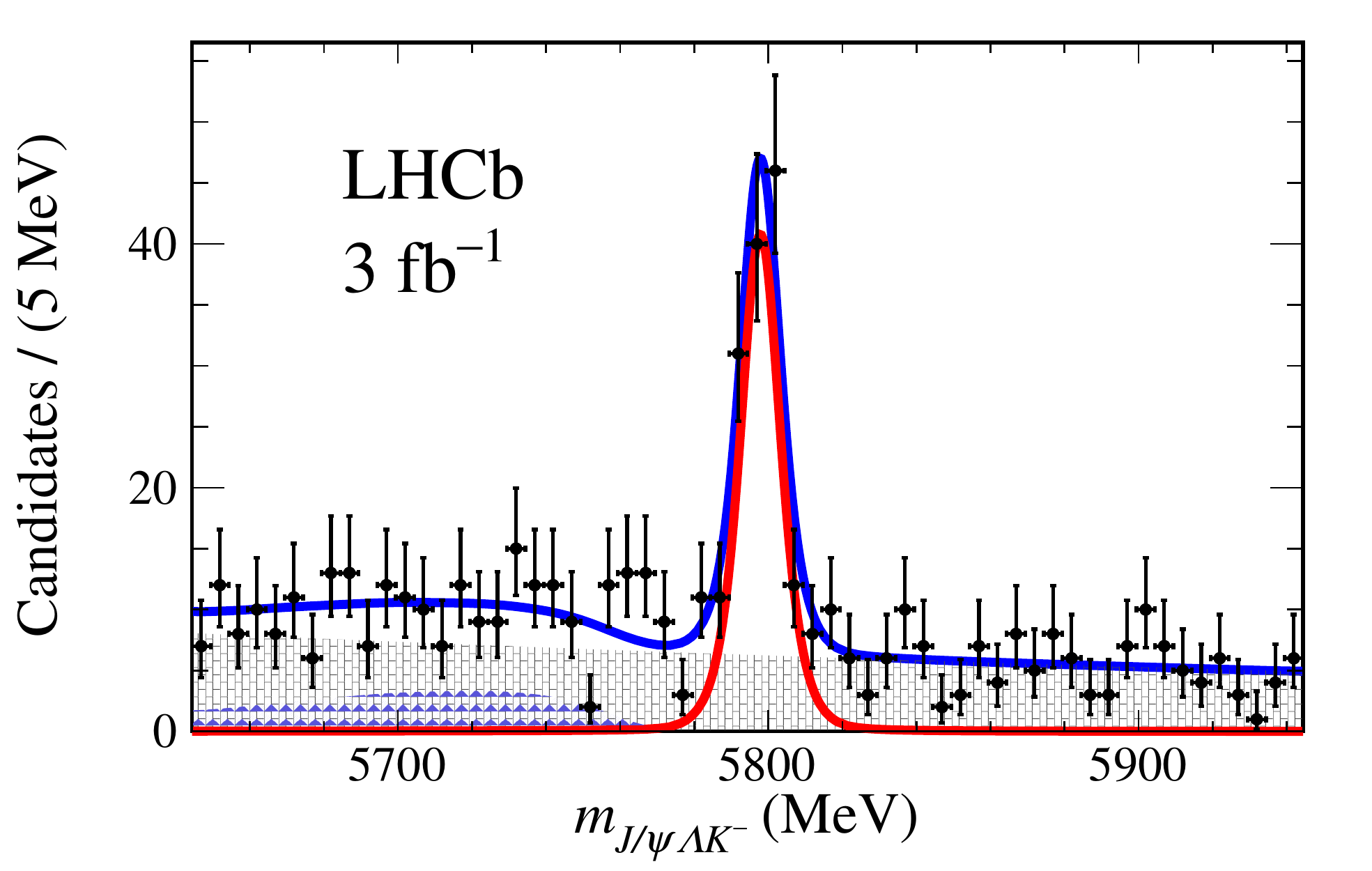}}\\
\sidesubfloat[]{
\includegraphics[width=0.45\textwidth]{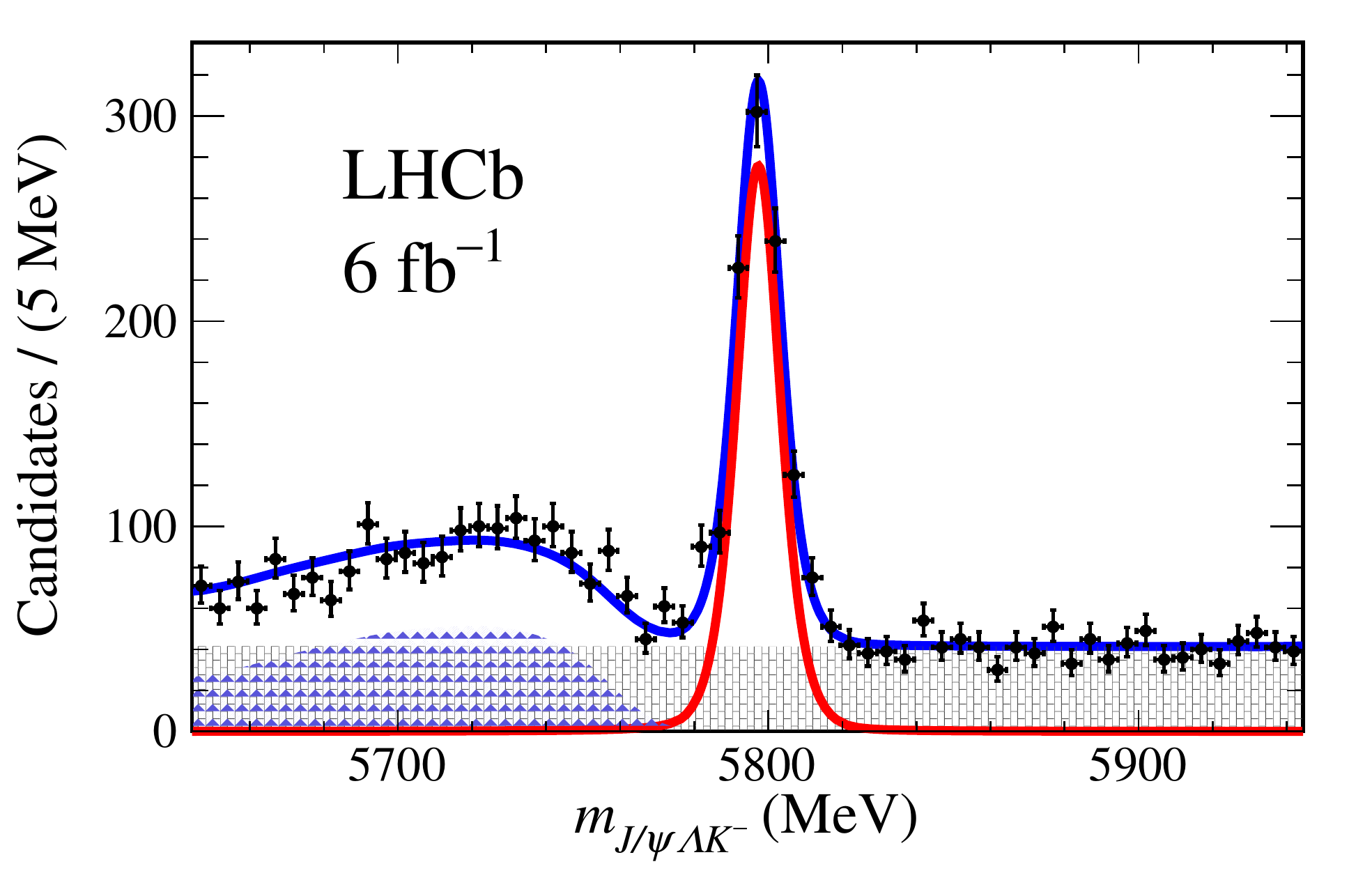}}%
\sidesubfloat[]{
\includegraphics[width=0.45\textwidth]{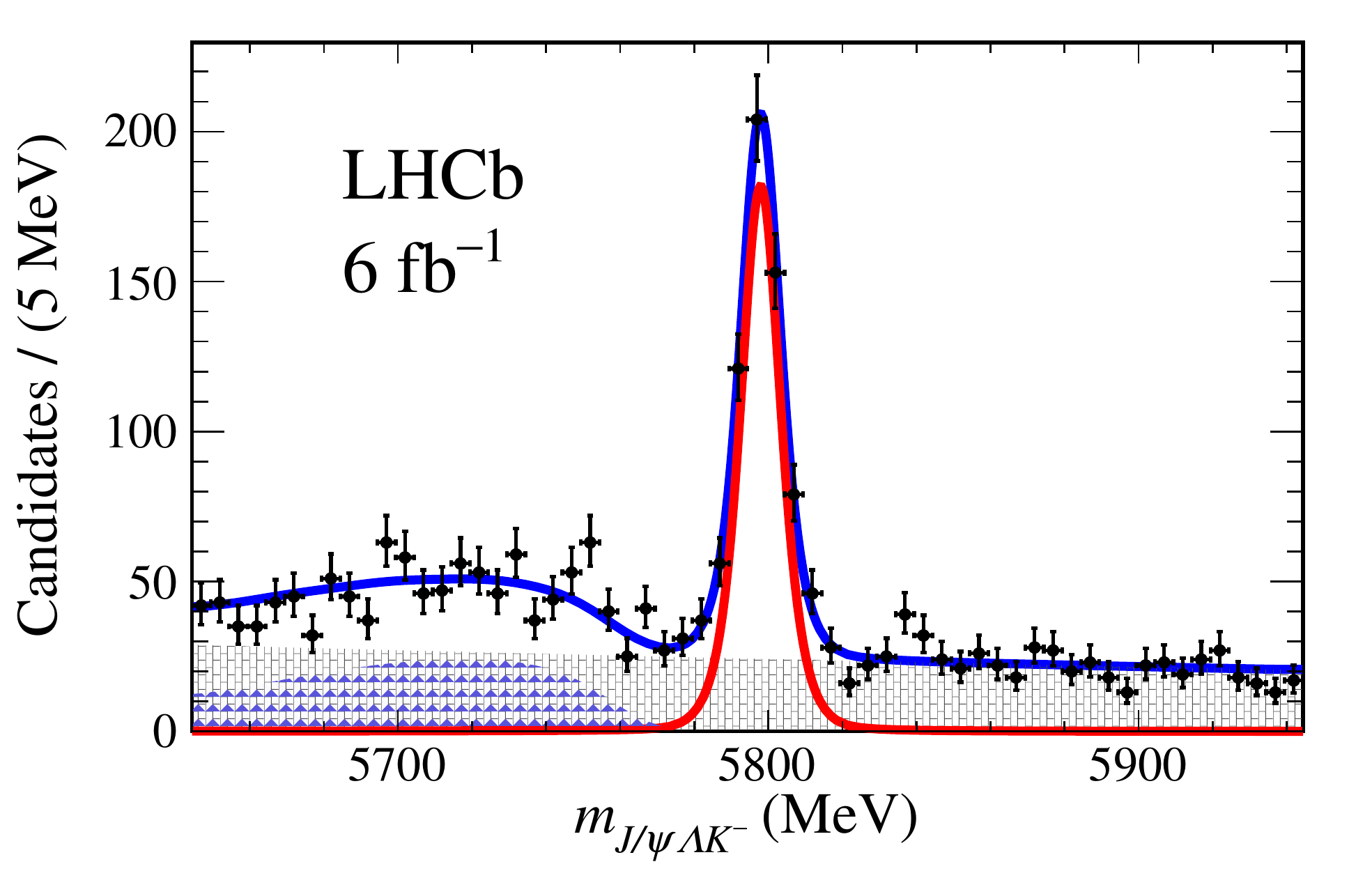}}
\caption{Invariant mass distributions of selected \XbJpsiLK candidates in the (a) Run 1 downstream, (b) Run 1 long, (c) Run 2 downstream, and (d) Run 2 long samples. The data are overlaid by the result of the fit described in the text.}
\label{massfit}
\end{figure}

Candidates are required to pass a set of selection criteria and are then further filtered using a multivariate classifier based on a gradient boosted decision tree (BDTG)~\cite{Breiman,Hocker:2007ht,TMVA4}. The selection criteria are almost identical to those used in the previous analysis~\cite{LHCb-PAPER-2016-053}, except those on the $\pt$ of the $\Lz$ decay products and on the $\chisqip$ of the kaon candidate which are relaxed. The $\chisqip$ is defined as the difference in the vertex-fit \chisq of a given PV reconstructed with and without the track considered. In total, 15 variables are combined to train the BDTG classifier. The requirement of the BDTG response is selected to maximise the signal significance, separately for four categories (long and downstream \Lz candidates in Run 1 and 2).  In the selected sample, less than 0.5\% of the events contain more than one \Xibm candidate, which are all retained.

\begin{figure}[b]
\centering
\includegraphics[width=0.65\textwidth]{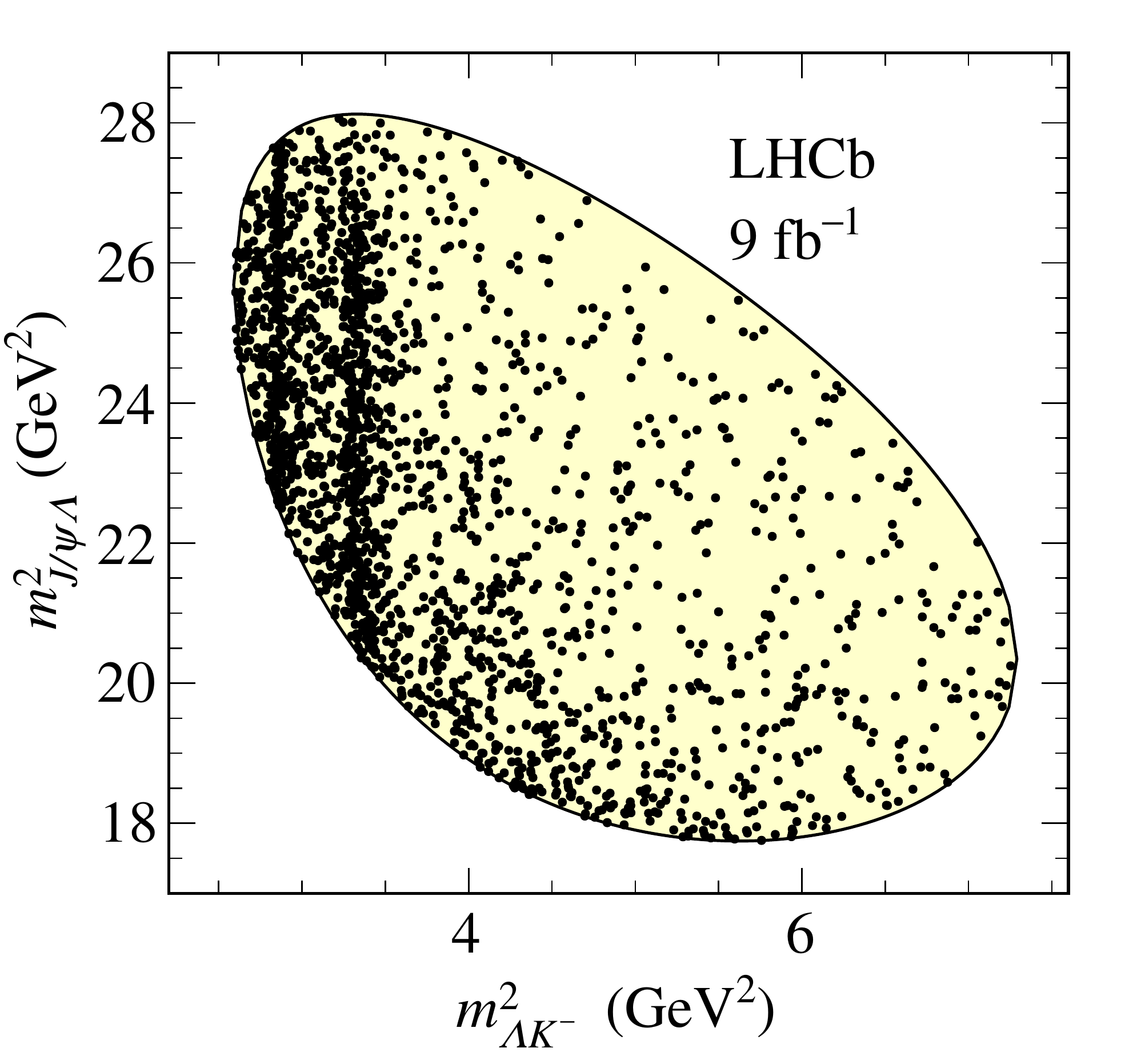}%
\caption{Dalitz plot for all candidates within $\pm15\mev$ of the known $\Xibm$ mass. The yellow area shows the kinematically allowed region.}
\label{dlz-point}
\end{figure}

A kinematic fit~\cite{Hulsbergen:2005pu} is applied to the $\Xibm$ decay to improve the mass resolution where the \jpsi and \Lz candidate masses are constrained to their known values~\cite{PDG2020}, and the $\Xibm$ candidate is constrained to originate from a primary vertex. The resulting $\jpsi\Lz\Km$ invariant mass spectra for the four categories are shown in Fig.~\ref{massfit}. 
The total signal yield is $1750\pm50$, determined by an unbinned extended maximum-likelihood fit to the \JpsiLK invariant mass spectra for each of the four categories. The signal is described by a Hypatia function~\cite{Santos:2013gra}, while combinatorial background is modelled by an exponential function. Partially reconstructed  \XbJpsiSK decays form a specific background at masses below the known \Xibm mass. The shape of this background is determined using a non-parametric model from simulation. Its yield varies freely in the fit. Weights~\cite{Pivk:2004ty} are assigned to the candidates to statistically subtract background contributions by weighting each candidate depending on their invariant \JpsiLK mass. A  kinematic fit is performed to improve the momentum resolution of final-state particles by further constraining the $\Xibm$ candidate mass to its known value~\cite{PDG2020}. The resulting Dalitz plot for candidates within $\pm15$\mev of the $\Xibm$ peak position is shown in Fig.~\ref{dlz-point}. As expected, significant $\Xistm\to \Lz \Km$ contributions, in particular from the \Xione and \Xitwo resonances, are observed. The $\JpsiL$ mass spectrum will be further explored in this article.

\section{Amplitude analysis}
An amplitude analysis is carried out to measure the properties of the \Xione and \Xitwo resonances, and to examine a possible contribution from any $\Pcs$ pentaquark states decaying into $\JpsiL$. The amplitude fits minimise an unbinned negative log-likelihood, $\Like$, constructed in a six-dimensional phase space of the data~\cite{LHCb-PAPER-2015-029}. The six dimensions correspond to the $\LK$ mass and five angular observables $\theta_{\Xibm}$\,, $\theta_{\Xim}$\,, $\theta_{\jpsi}$\,, $\phi_{\Lz}$\,, $\phi_{\mu}$\,,
where $\theta$ and $\phi$ denote the polar and azimuthal angles, respectively. The probability distribution function only comprises the contribution from signal \Xibm decays, since the background is subtracted using the sPlot technique~\cite{Pivk:2004ty,Xie:2009rka}, as discussed in detail in Ref.\cite{LHCb-PAPER-2015-029}. The efficiency on the six-dimensional phase space is folded into the signal probability density function. The  amplitude analysis follows a similar strategy to that of $\LbJpsipK$ decays in  Ref.~\cite{LHCb-PAPER-2015-029}, with the $\Lb$ baryon and proton in the $\Lb$ decay replaced by the $\Xibm$ and $\Lz$ baryons, respectively. However, a cross-check with the  Dalitz-plot decomposition method proposed in Ref.~\cite{Mikhasenko:2019rjf} indicates that the method used in  Ref.~\cite{LHCb-PAPER-2015-029} has to be modified in two aspects to properly align the helicity state of the spin-half \Lz baryon in the $\Xistm$ and $\Pcs$ decay chains~\cite{Wang:2020giv}. First, in the $\Xistm\to\Lz\Km$ decay, the $\Lz$ particle is used to define the two decay angles of the $\Xistm$ system. The definition of the remaining angles is the same as in  Ref.~\cite{LHCb-PAPER-2015-029}. Secondly, the Euler rotation in the $\Xistm\to\Lz\Km$ decay aligns the spin axis along the \Lz momentum, while the rotation in the $\Pcs\to\jpsi\Lz$ frame aligns the spin axis in the direction opposite to the \Lz momentum. An additional rotation to align the $z$-axis between the $\Pcs$ and $\Xistm$ chains generates a term $(-1)^{J_\Lz-\lambda^{\pcs}_\Lz}$ in the amplitude of the $\Pcs$ chain, where $J_\Lz$ and $\lambda^{\pcs}_\Lz$ are the spin and the helicity of the $\Lz$ particle in the $\Pcs$ rest frame, respectively. This term is the particle-two convention factor~\cite{Jacob:1959at}. 

\begin{table}[b]
\centering
\caption{The components in the amplitude fit used to describe the $\LK$ system. 
The $J^P$, masses ($M_0$) and widths ($\Gamma_0$) of the $\PXi^-$  states are taken from the PDG~\cite{PDG2020}. The numbers of $LS$ couplings used in the default fit are listed, together with the total number of the $LS$ couplings associated to the $\LK$ component, given in parentheses. The $\PXi(1820)^-$ coupling of lowest $LS$ is set to (1,0) for reference. Multiple $J^P$ assignments are considered for states where this assignment has not been previously established. A nonresonant $S$-wave \LK contribution, labelled as NR, is also considered in the fit model.}
\label{tab:Xistar}
\vspace{0.2cm}
\begin{tabular}{llccc}
\hline\\[-2.5ex] 
State & $M_0$ (MeV) & $\Gamma_0$ (MeV)&  $LS$ couplings & $J^P$ examined \\
\hline \\[-2.5ex] 
$\PXi(1690)^-$ & $1690\pm10$ & $<30$ & 4 (6) &$(1/2, 3/2)^\pm$ \\
$\PXi(1820)^-$ & $1823\pm5$ & $24_{-10}^{+15}$& 3 (6) & $3/2^-$\\
$\PXi(1950)^-$ &$1950\pm15$ & $60\pm20$ & 3 (6)  & $(1/2, 3/2, 5/2)^\pm$\\
$\PXi(2030)^-$ &  $2025\pm5$ & $20_{-5}^{+15}$ & 3 (6) & $5/2^\pm$ \\
NR $\LK$   &  \multicolumn{1}{c}{-} & - & 4 (4) & $1/2^-$  \\
\hline
\end{tabular}
\end{table}

\begin{figure}[b]
\centering
\sidesubfloat[]{
\includegraphics[width=0.45\textwidth]{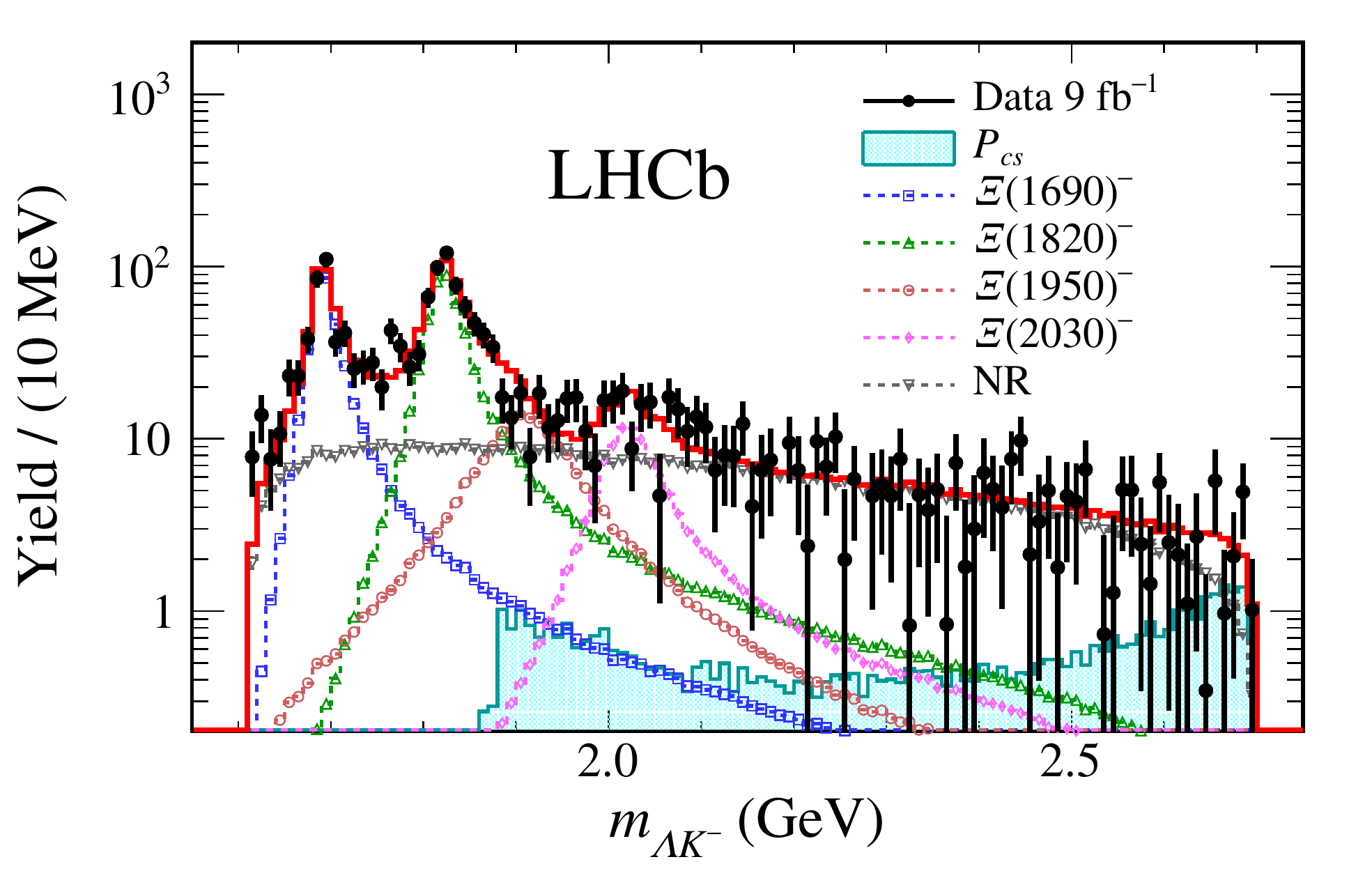}}%
\sidesubfloat[]{
\includegraphics[width=0.45\textwidth]{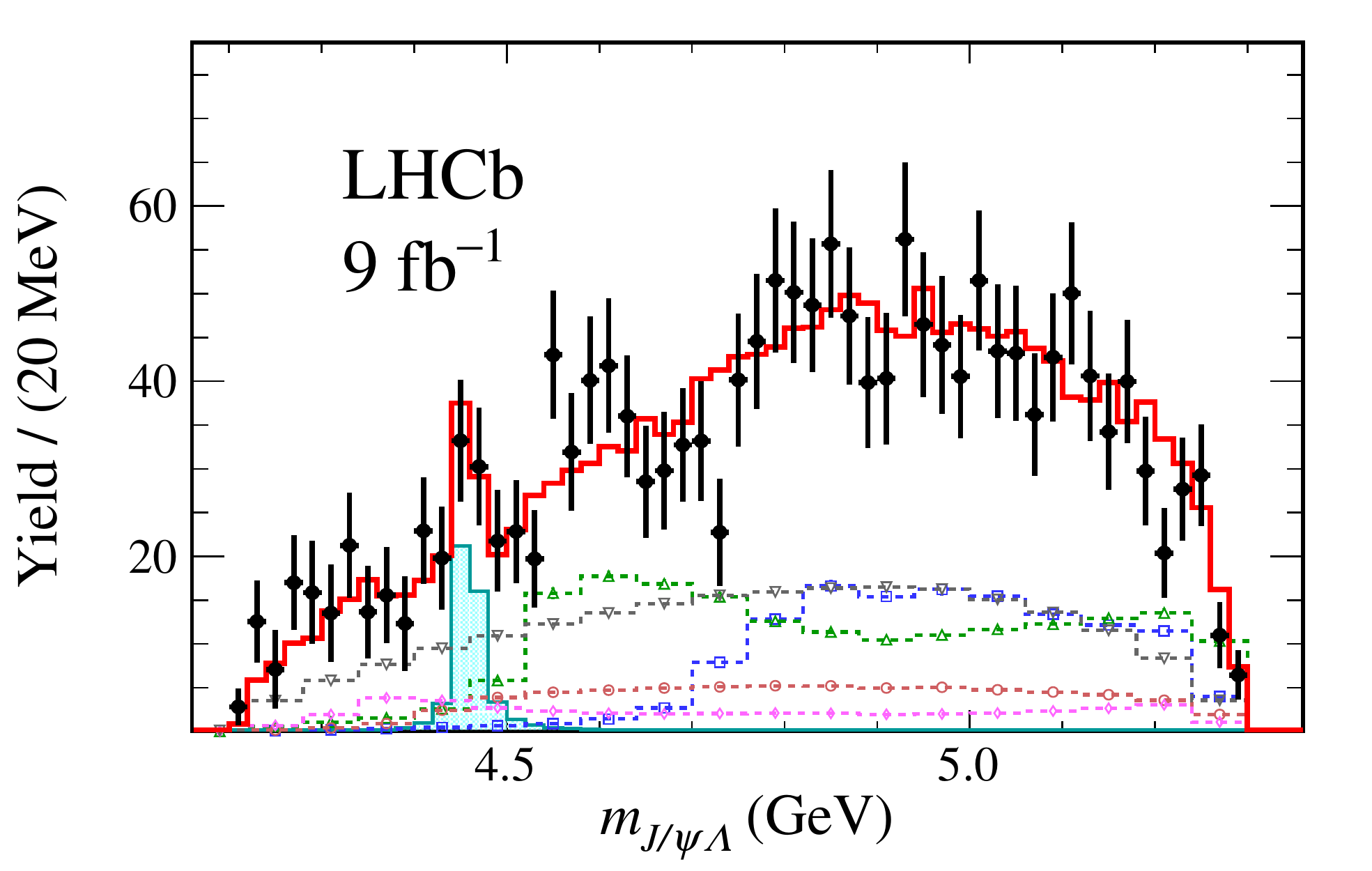}}
\caption{The (a) $m_{\LK}$ and (b) $m_{\jpsi\Lz}$ distributions of selected candidates compared to the result of the fit with the \Pcs state. 
}
\label{fig:fit}
\end{figure}

Table~\ref{tab:Xistar} lists the possible contributions from well established \Xistm states according to the PDG~\cite{PDG2020}. The states constitute a default description of the $\LK$ invariant mass spectrum, which is also assumed to include a nonresonant (NR) contribution.
As the spin-parities of these resonances are unknown except for that of the \Xitwo baryon, combinations of different $J^P$ of these states are examined in the amplitude fit.
Due to limited sample size, each \Xistm resonance is described by 3 or 4 independent $LS$ couplings, where $L$ stands for the decay orbital angular momentum, and $S$ is the sum of spins of the decay products.   Couplings corresponding to higher $L$ are expected to be suppressed by the angular momentum barrier, so the contributions are chosen in increasing order of $L$. For all \Xistm resonances, relativistic Breit-Wigner functions~\cite{LHCb-PAPER-2015-029} are used to model their line shape and phase variation as a function of the invariant mass of the \LK system, $m_{\LK}$. The masses and widths of the \Xione and \Xitwo resonances are free fit parameters, 
while those of other \Xistm resonances are constrained by their known uncertainties~\cite{PDG2020}. The contribution of the NR $S$-wave component to the $m_{\LK}$ spectrum is described with a function that is inversely proportional to $m_{\LK}^2$~\cite{Cesar:2016kns}. Alternative descriptions of the NR component are considered to estimate systematic uncertainties on the model. The $\LK$ and $\JpsiL$ mass spectra are shown in Fig.~\ref{fig:fit} with the projections of amplitude fit overlaid.

After the determination of the amplitude model with $\LK$-only contributions, a \Pcs state is added to the amplitude model, with spin hypotheses ranging from $1/2$ to $5/2$ and parity hypotheses of both $-1$ and $+1$. Only the smallest allowed $L$ is considered due to the suppression of higher values of $L$. A \JpsiL mass resolution of 2.6\mev obtained from simulation is taken into account by smearing the \Pcs Breit-Wigner amplitude accordingly. The fits show a significant improvement when adding the \Pcs state. 
The largest improvement on $-2\ln\Like$ when adding a single \Pcs contribution is found to be $\Delta2\ln\Like = 43$, for an addition of 6 parameters. 
This fit, which includes the $\LK$ resonances in Table~\ref{tab:Xistar}, a NR $\LK$ component and a single $\Pcs$ state each with their favoured $J^P$ assignment is referred to below as the default fit. 
The improvement in $-2\ln\Like$
corresponds to a statistical significance of $4.3$ standard deviations ($\sigma$).
This is estimated using pseudoexperiments where the look-elsewhere effect is taken into account. The difference of the $-2\ln\Like$ obtained using fit models with and without the contribution of the $\Pcs$ state is used as the test statistic to evaluate the $p$-value of the null hypothesis, where several alternative $\Xistm$ models are used to describe the contributions from the $\LK$ resonances. 
The $p$-value is estimated by fitting the distribution of the test statistic from 10\,000  pseudoexperiments for the model based on the results from the fit to data fit, generated with the null hypothesis. 
To take into account the look-elsewhere effect for each pseudoexperiment, the global maximum of $2\ln\Like$ is obtained by scanning the values of the mass and width of $\Pcs$ state in the kinematically allowed region, instead of limiting their values to be consistent with that of the data fit. When including systematic uncertainties discussed below, the $p$-value is determined to be 0.2\% by counting the fraction of pseudoexperiments with the $\Delta2\ln\Like$ value exceeding the smallest $\Delta2\ln\Like$ value from data. This $p$-value corresponds to the signal significance of $3.1\sigma$ with a two-sided Gaussian test for the $\Pcs$ state, providing the first evidence for a charmonium pentaquark candidate with strangeness.

\begin{figure}[t]
\centering
\sidesubfloat[]{
\includegraphics[width=0.45\textwidth]{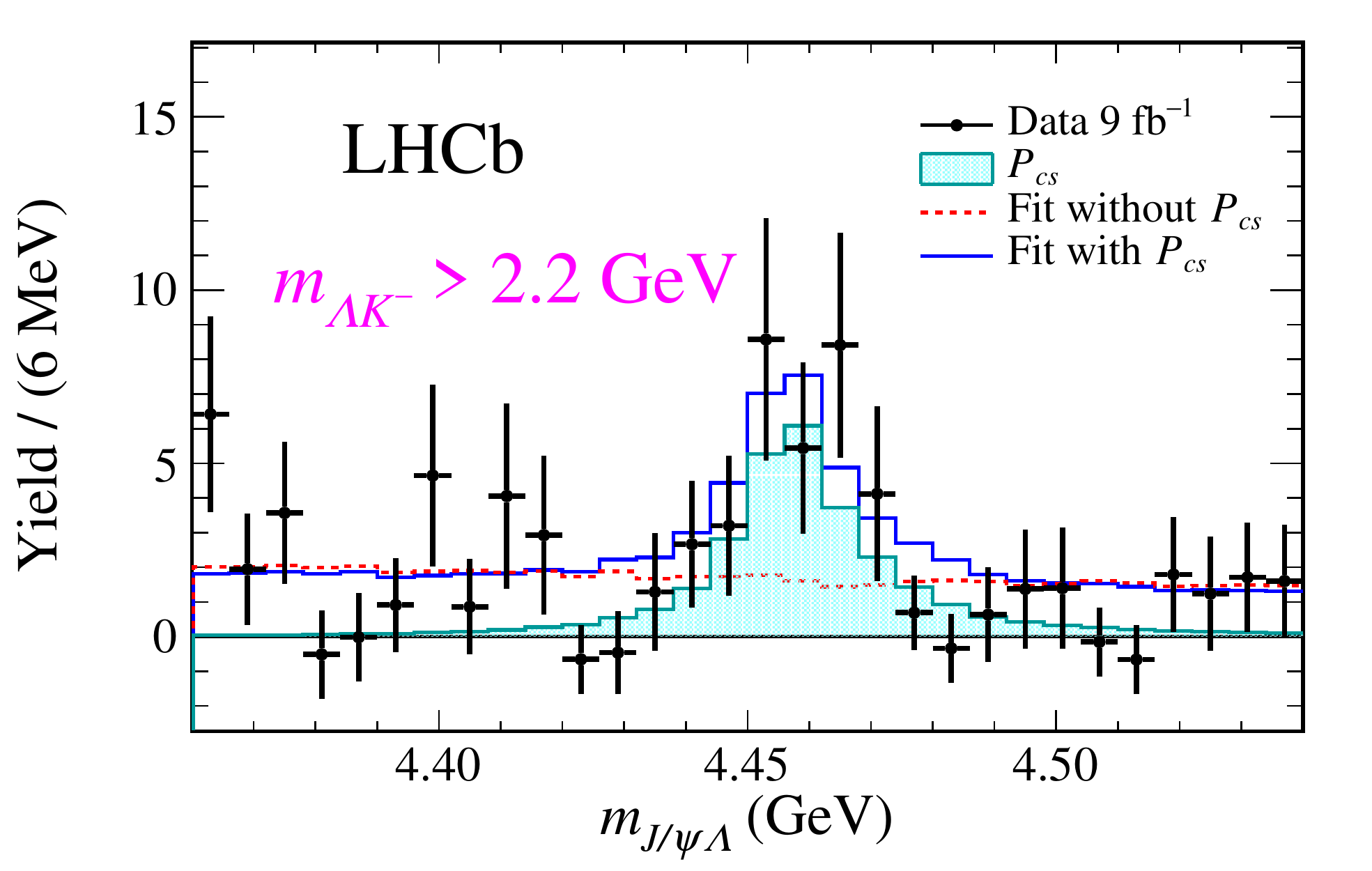}}%
\sidesubfloat[]{
\includegraphics[width=0.45\textwidth]{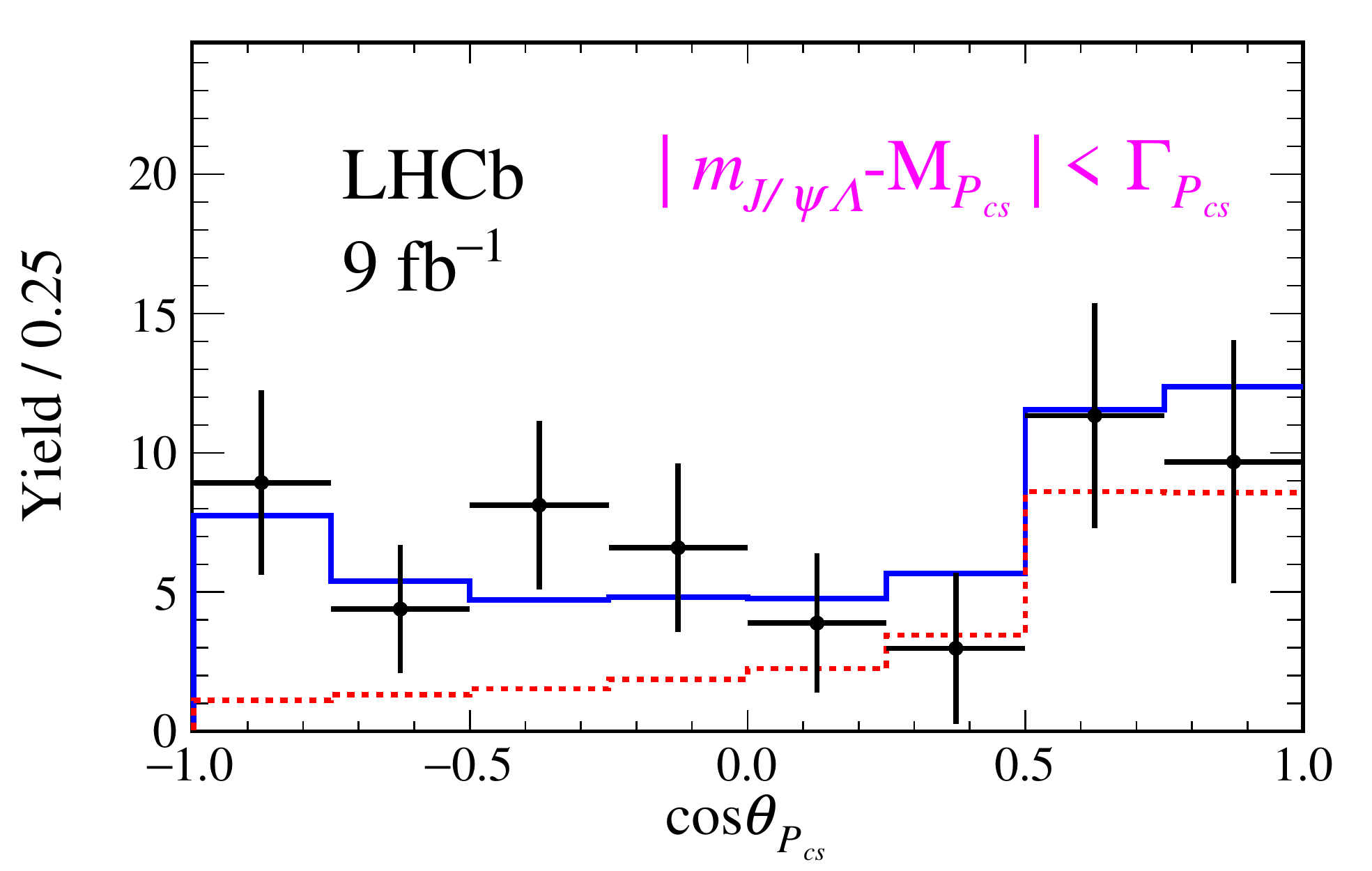}}
\caption{(a) Projection of $\mPc$ in the range of $[4.36, 4.54]\gev$ with the $\m(\LK)>2.2$\gev requirement. (b) Projection of $\cos\theta_{P_{cs}}$  for candidates having \mPc within one natural width from the fitted $\Pcs$ resonance mass. The red dashed lines show the result of the fit in $\PXi^{*}$-only hypothesis and the blue solid lines show the fit accounting for both $\PXi^{*}$ and $\Pcs$ states.  }
\label{fig:imp1z}
\end{figure}

\begin{figure}[b]
\centering
\includegraphics[width=0.8\textwidth]{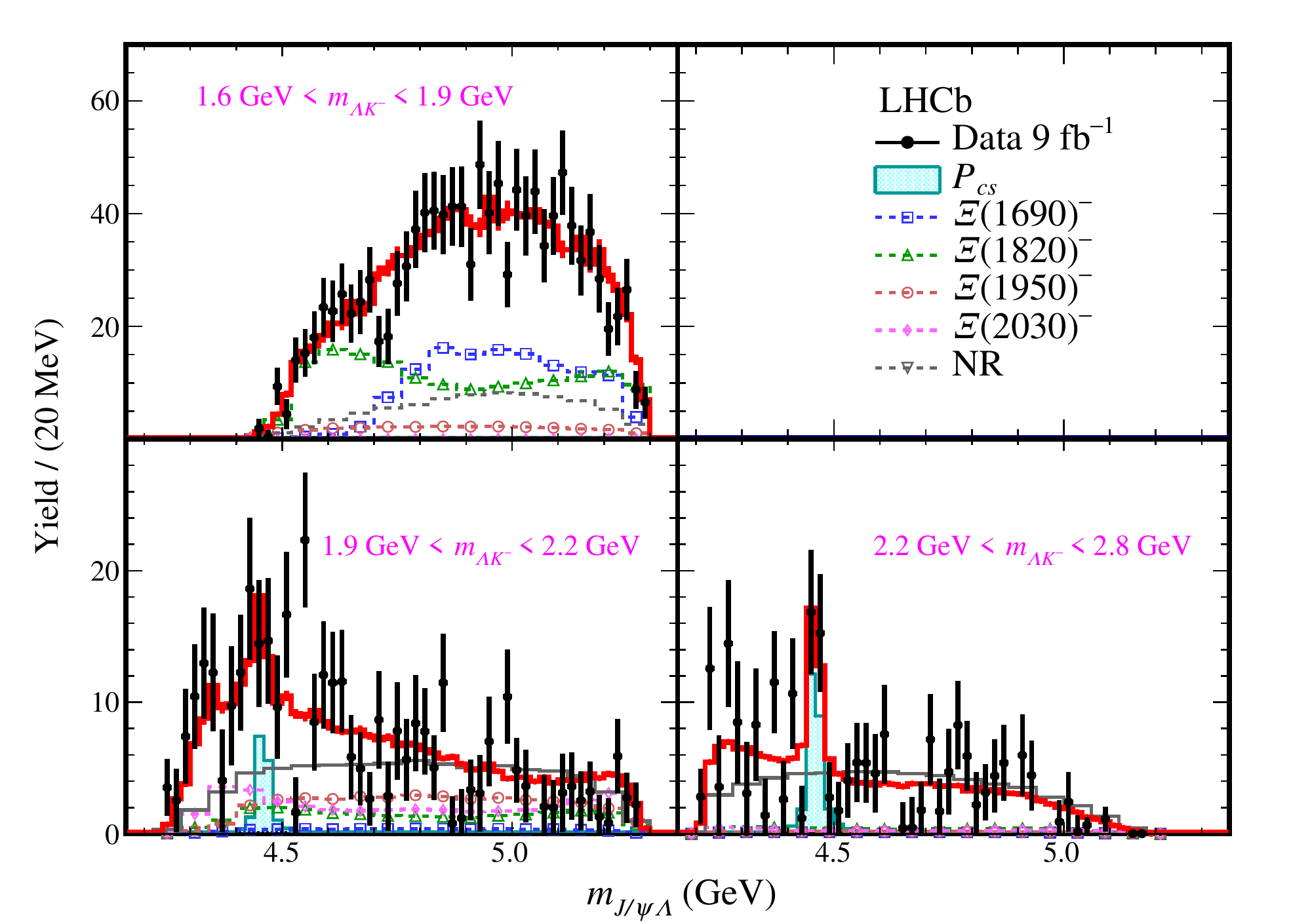}%
\caption{Projections of $m_{\JpsiL}$ in intervals of (top left) $m_{\LK}<1.9$\gev  (bottom left) $1.9<m_{\LK}<2.2$\gev and (bottom right) $m_{\LK}>2.2\gev$ based on the default fit, superimposed with contributions from components listed in Table~\ref{tab:Xistar} and the \Pcs state.
}
\label{fig:3bin}
\end{figure}

As shown in Fig.~\ref{fig:fit}, the projections of the full amplitude fit onto the \LK and \JpsiL invariant mass spectra match the data distributions well. A test of the fit quality is performed by comparing the default fit of the Dalitz plot with the data distribution. The data is divided into 64 bins containing approximately the same number of decays. The $\chisq$ is calculated to be 77 for these bins, indicating a reasonably good description of the data. The \Pcs state is determined to have a mass of $4458.8\pm2.9 \stat \mev$ and a width of $17.3\pm6.5\stat\mev$, and hereafter is denoted as \TPcs.  Figure~\ref{fig:imp1z} highlights the \TPcs contribution by comparing the fits to the $m_{\JpsiL}$ and $\cos\theta_{P_{cs}}$ distributions with and without the \TPcs state included, where $\theta_{\pcs}$ is the helicity angle of the $\JpsiL$ system, defined as the angle between the direction of the $\jpsi$ particle and the opposite direction of the $K^-$ particle in the $\JpsiL$ rest frame. The \TPcs state is more visible when the dominant contributions from $\Xistm$ with low masses are suppressed by requiring $m_{\Lz\Km} > 2.2\gev$. As shown in Fig.~\ref{fig:imp1z} (right), a significant improvement of the fit quality is also found in the $\cos\theta_{\pcs}$ distribution when including the $\TPcs$ state. 

No evidence for any other \Pcs state is found in the considered mass range. This is also clear when examining the $m_{\JpsiL}$ projections in three intervals of $m_{\LK}$, shown in Fig.~\ref{fig:3bin}. The measured mass, width and fit fraction (FF) of all components involved in the default fit are shown in Table~\ref{tab:final}. Systematic uncertainties on these results are discussed below.

\begin{table}[t]
\centering
\caption{Mass ($M_0$), width ($\Gamma_0$) and fit fraction (FF) of the components involved in the default fit.  The masses and widths of the \Pcs, \Xione, and \Xitwo resonances are free parameters, while those of the other \Xistm resonances are constrained by the known uncertainties~\cite{PDG2020}. The quoted uncertainties are statistical and systematic. When only one uncertainty is given, it is statistical. }\label{tab:final}
\def\arraystretch{1.2}
\begin{tabular}{clll}\hline
State&$M_0$ (MeV)&$\Gamma_0$ (MeV)& FF (\%)\\\hline
$\TPcs$&$	4458.8 	\pm	2.9 \,	_{-	1.1 	}^{+	4.7 	}$&$	17.3 	\pm	6.5 \,	_{-\,	5.7 	}^{+\,	8.0 	}$&$	2.7 \,	_{-\,	0.6 	\, -\,	1.3 	}^{+\,	1.9 	\, +\,	0.7 	}$\\
$\PXi(1690)^-$&$	1692.0 	\pm	1.3 \,	_{-\,	0.4 	}^{+\,	1.2 	}$&$	25.9 	\pm	9.5 \,	_{-\,	13.5 	}^{+\,	14.0 	}$&\hspace{-0.5em}$	22.1 \,	_{-\,	2.6 \,	-\,	8.9 	}^{+\,	6.2 \,	+\,	6.7 	}$\\
$\PXi(1820)^-$&$	1822.7 	\pm	1.5 \,	_{-\,	0.6 	}^{+\,	1.0 	}$&$	36.0 	\pm	4.4 \,	_{-\,	8.2 	}^{+\,	7.8 	}$&\hspace{-0.5em}$	32.9 \,	_{-\,	6.2 \,	-\,	4.1 	}^{+\,	3.2 \,	+\,	6.9 	}$\\
$\PXi(1950)^-$&$    1910.6  \pm 18.4                $&\hspace{-0.5em}$      105.7 \pm 23.2         $&\hspace{-0.5em}$ 11.5 \, _{-\,3.5 \, -\,\phantom{0}9.4} ^{+\,5.8 \, +\,49.9}$\\
$\PXi(2030)^-$&$    2022.8  \pm 4.7                $&$      68.2  \pm 8.5          $&$ 7.3\, _{-\,1.8\,-\, 4.1 }^{+\, 1.8\,+\,3.8 }$\\
 NR     &$    -               $&$         -           $&\hspace{-0.5em}$     35.8 \, _{-\,6.4 \, -11.2}   ^{+\,4.6 \, +10.3}    $\\ 
\hline
\end{tabular}
\end{table}

\begin{figure}[b]
\centering
\includegraphics[width=0.5\textwidth]{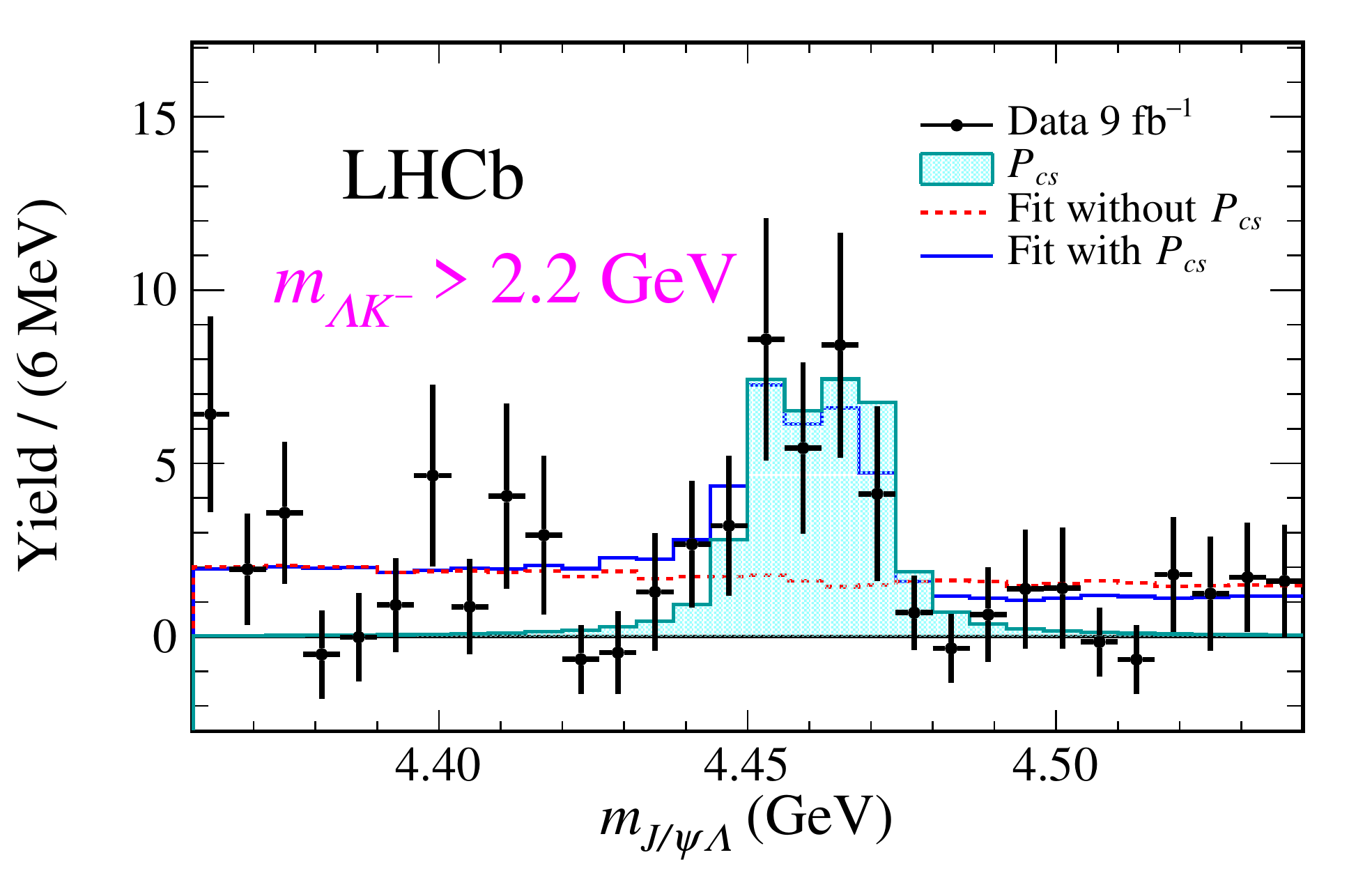}%
\caption{Projection of $\mPc$ with a $\mLK>2.2$\gev requirement applied, overlaid by the fit using two resonances to model the peak region.}
\label{fig:2peak}
\end{figure}

The measured $\TPcs$ mass is about 19\mev below the $\PXi_c^0\Dstarzb$ threshold. In this region, Ref.~\cite{Wang:2019nvm} predicts two states with $J^P=1/2^-$ and $3/2^-$ and a mass difference of 6\mev. This is similar to the two $P_c(4440)^+$ and $P_c(4457)^+$ pentaquark states, which are just below the $\PSigma_c^+\Dstarzb$ threshold. Thus the hypothesis of a two-peak structure with the predicted $J^P$ values is tested. The fit provides a good description of the data. The result is shown for the $\Pcs$ enhanced region in Fig.~\ref{fig:2peak}. The masses and widths of the two states are $4454.9\pm2.7$\mev and $7.5\pm9.7$\mev, and $4467.8\pm3.7$\mev and $5.2\pm5.3$\mev, respectively, where the uncertainties are statistical only. The fit improves $2\ln\Like$ by 4.8 units for 4 additional free parameters, compared to the fit using one Breit-Wigner function to model the structure. Therefore, the analysis of the current data sample cannot confirm or refute the two-peak hypothesis.

Systematic uncertainties are summarised in Table~\ref{tab:sys}. 
Due to the limited number of candidates, $J^P$ for the \Pcs state and for the \Xione, \Xithree and \Xifour states cannot be reliably determined. 
To estimate the systematic uncertainty related to the choice of $J^P$ assignments, all possible $J^P$ assignments for these states resulting in $-2\ln\Like$ differences less than 9 units are compared to that from the default fit. 
The largest variation, with respect to the values in the default fit, is taken as an asymmetric systematic uncertainty. The uncertainty related to the fit model is estimated by: varying the hadron-size parameter in the Blatt-Weisskopf barrier factor~\cite{LHCb-PAPER-2015-029}  between 1.0 and \mbox{$5.0 \gev^{-1}$}; modifying the orbital angular momenta $L$ in \Xibm decays that are used in the treatment of the \Xistm resonances by one or two units; using all allowed couplings, instead of a limited number of $L$ couplings, for \Xistm or \TPcs states; using alternative models to describe the  nonresonant \LK component,  including an exponential function or a function inversely proportional to $m^2_{\LK}+m^2_{\rm{NR}}$, where $m_{\rm{NR}}$ is a free parameter in the fit; considering the effects of \Xibm polarisation, which are found to be consistent with zero in the analysis and neglected in the default fit; using an extended $\Xistm$ model, which includes two more $\Xistm$ states, in which the mass and width constraints on the \Xistm states are removed, and all allowed couplings for all \Xistm states are used. The largest value among all model variations is taken as systematic uncertainty for this source. The other systematic sources are estimated by:  including the $\Lz \to p\pim$ decay angles instead of taking the $\Lz$ baryon as a stable final-state particle; determining the sWeights by either splitting \Xistm helicity angles into bins or removing partially reconstructed physical background from the \XbJpsiSK decays in the [5644.5,5764.5]\mev \JpsiLK mass sideband; and varying the efficiency due to imperfect simulation. The mass resolution of the \LK system is about 1-2 \mev, and has negligible effect on the fit due to the large widths of the \Xistm states. The significance for the \TPcs state is conservatively taken as the smallest significance found when combining different sources of systematic uncertainty together, and is equal to 3.1$\sigma$, as already reported, where the look-elsewhere effect has been taken into account. It corresponds to varying the hadron-size parameter in the extended  $\Xistm$ model with full couplings for the considered \TPcs state.

The negative systematic uncertainty for the $\TPcs$ fit fraction, $-1.3\%$, is obtained from an alternative value of $J^P$ used for the $\TPcs$ state. In such a fit, the significance of the $\TPcs$ state is 4.1$\sigma$, even though the fit fraction is $1.4\%$. This is because the significance has contributions from two sources, the fit fraction and the interference fraction involving the $\TPcs$ state. The interference fraction is about $+1.3\%$ in this alternative $J^P$ fit, while it is almost 0 in the default fit. 
The systematic uncertainty of the \Xithree fit fraction is $+49.9\%$, most of which originates from an alternative fit where its mass and width are floated in the extended model, rather than constrained to the known values~\cite{PDG2020}, while the second largest one, from other sources considered in the estimation of systematic uncertainty, is $+5.9\%$. Considering this large value, the fit fractions for all components involved in the extended model and their interference fractions are checked. A large interference fraction of $-$60.3\% between \Xithree and NR is found in the extended model, and a large width of $\Xithree$ of about 350\mev is found. This could indicate the NR description in the default model is not perfect. Therefore several other NR models discussed before, or the NR contribution replaced by a broad Breit-Wigner function are tested; all of these variations don't change the conclusion on the $\TPcs$ result.

\begin{table}[t]
\centering
\caption{Summary of absolute systematic uncertainties for the fit parameters. The units for masses ($M_0$) and widths ($\Gamma_0$) are $\mev$. The fit fraction in percent is denoted FF.
}\label{tab:sys}
\def\arraystretch{1.2}
\begin{tabular}{lcccccccccp{10mm}p{10mm}c}\hline
\vspace{-2mm}
Source & \multicolumn{3}{c}{$\TPcs$}&\multicolumn{3}{c}{$\Xione$}&\multicolumn{3}{c}{$\Xitwo$}&$\Xistm$&$\Xistm$& $NR$\\
 & \multicolumn{9}{c}{}&$(1950)$&$(2030)$& \\
  & $M_0$ &$\Gamma_0$& FF & $M_0$ &$\Gamma_0$& FF & $M_0$ &$\Gamma_0$& FF & FF & FF & FF\\\hline
$J^P$	& $_{-	0.3 	}^{+	4.7 	}$&	$_{-	5.7 	}^{+	0.0 	}$&	$_{-	1.3 	}^{+	0.1 	}$ &$_{-	0.1 	}^{+	1.2 	}$ &$_{-	\phantom{0}0.9 	}^{+	14.0 	}$ &$_{-	0.3 	}^{+	6.7 	}$ &$_{-	0.2 	}^{+	0.8 	}$ &$_{-	0.5 	}^{+	1.4 	}$ &$_{-	0.3 	}^{+	4.2 	}$ & $_{- \phantom{0}9.4 }^{+\phantom{0} 0.2 }$ & $_{- 4.1 }^{+ 0.0 }$ & $_{- 11.2 }^{+ \phantom{0}0.9 }$	\\
Model	&$_{-	1.1 	}^{+	0.7 	}$ &$_{-	2.0 	}^{+	8.0 	}$	&$_{-	0.5 	}^{+	0.7 	}$ &$_{-	0.4 	}^{+	0.5 	}$ &$_{-	13.5 	}^{+	\phantom{0}1.8 	}$ &$_{-	8.9 	}^{+	1.9 	}$ &$_{-	0.6 	}^{+	1.0 	}$ &$_{-	8.2 	}^{+	7.8 	}$ &$_{-	4.1 	}^{+	6.9 	}$ & $_{- \phantom{0}5.4 }^{+ 49.9 }$ & $_{- 1.6 }^{+ 3.8 }$ & $_{- \phantom{0}6.4 }^{+ 10.3 }$\\
$\Lz$ decay &$_{-	0.7 	}^{+	0.0 	}$ &$_{-	4.7 	}^{+	0.0 	}$	& $_{-	0.3 	}^{+	0.0 	}$&$_{-	0.4 	}^{+	0.0 	}$ &$_{-	\phantom{0}0.0 	}^{+\phantom{0}	0.2 	}$ &$_{-	0.8 	}^{+	0.0 	}$ &$_{-	0.5 	}^{+	0.0 	}$ &$_{-	7.2 	}^{+	0.0 	}$ &$_{-	4.1 	}^{+	0.0 	}$ & $_{- \phantom{0}0.0 }^{+ \phantom{0}2.4 }$ & $_{- 1.3 }^{+ 0.0 }$ & $_{- \phantom{0}0.0 }^{+ \phantom{0}3.9 }$\\
sWeights	&$_{-	0.2 	}^{+	0.0 	}$ &	$_{-	0.0 	}^{+	0.3 	}$	&$_{-	0.0 	}^{+	0.1 	}$ &$_{-	0.1 	}^{+	0.1 	}$ &$_{-\phantom{0}	0.2 	}^{+\phantom{0}	3.1 	}$ & $_{-	0.0 	}^{+	1.4 	}$&$_{-	0.2 	}^{+	0.2 	}$ & $_{-	1.5 	}^{+	2.2 	}$&$_{-	0.5 	}^{+	1.6 	}$ & $_{- \phantom{0}1.6 }^{+ \phantom{0}0.7 }$ & $_{- 0.2 }^{+ 0.0 }$ & $_{- \phantom{0}2.7 }^{+ \phantom{0}0.0 }$\\
Efficiency	&$_{-	0.1 	}^{+	0.1 	}$ &$_{-	0.5 	}^{+	0.0 	}$	&$_{-	0.1 	}^{+	0.0 	}$&$_{-	0.2 	}^{+	0.1 	}$ &$_{-\phantom{0}	1.5 	}^{+\phantom{0}	2.1 	}$ &$_{-	1.3 	}^{+	0.8 	}$ &$_{-	0.2 	}^{+	0.1 	}$ &$_{-	0.3 	}^{+	1.1 	}$ &$_{-	0.7 	}^{+	0.5 	}$ & $_{- \phantom{0}1.0 }^{+ \phantom{0}2.3 }$ & $_{- 0.2 }^{+ 0.3 }$ & $_{- \phantom{0}0.9 }^{+ \phantom{0}1.1 }$\\
\hline
Final	&$_{-	1.1 	}^{+	4.7 	}$&$_{-	5.7 	}^{+	8.0 	}$	&$_{-	1.3 	}^{+	0.7 	}$ &$_{-	0.4 	}^{+	1.2 	}$ &	$_{-	13.5 	}^{+	14.0 	}$ &$_{-	8.9 	}^{+	6.7 	}$ &$_{-	0.6 	}^{+	1.0 	}$ & $_{-	8.2 	}^{+	7.8 	}$&$_{-	4.1 	}^{+	6.9 	}$ & $_{- \phantom{0}9.4 }^{+ 49.9 }$ & $_{- 4.1 }^{+ 3.8 }$ & $_{- 11.2 }^{+ 10.3 }$\\\hline
\end{tabular}
\end{table}

\section{Conclusion}

In conclusion, an amplitude analysis of \XbJpsiLK decays is performed using approximately 1750 candidates, and a structure in the \JpsiL mass spectrum around 4459\mev is seen. This structure can be explained by including a pentaquark candidate with strangeness in the amplitude model. Its significance exceeds 3$\sigma$ after considering all systematic uncertainties. The mass and width of this new exotic state are measured to be $4458.8\pm\,2.9^{\,+\,4.7}_{\,-\,1.1}\mev$ and $17.3\pm\,6.5^{\,+\,8.0}_{\,-\,5.7}\mev$, respectively. The  \TPcs state has a mass only about 19\mev below the $\PXi_c^0\Dstarzb$ threshold and a narrow width. Motivated by this fact, the hypothesis of two resonances contributing to the enhancement is tested. The data cannot confirm or refute the two-peak hypothesis.  Furthermore, two \Xistm states, \Xione and \Xitwo, are observed for the first time in $\Xibm$ decays. Using the full amplitude analysis, their masses and widths are measured to be
\begin{equation*}
\begin{aligned}
   & M(\Xione) = 1692.0\pm1.3\,^{+\,1.2}_{-\,0.4} \mev,\, & \Gamma(\Xione) = 25.9\pm9.5\,^{+\,14.0}_{-\,13.5} \mev, \\
   & M(\Xitwo) = 1822.7\pm1.5\,^{+\,1.0}_{-\,0.6} \mev,\, & \Gamma(\Xitwo) = 36.0\pm4.4\,^{+\,7.8}_{-\,8.2} \mev.
\end{aligned}
\end{equation*}
These results are consistent with the average values reported in the PDG~\cite{PDG2020} and recent results from the BESIII experiment~\cite{Ablikim:2015apm,Ablikim:2019kkp}. The mass determinations are much more precise than those listed in the PDG. Due to limited signal yield, the  $J^P$ of the \TPcs and \Xione states are not determined at this stage.

\section*{Acknowledgements}
\noindent The numerical calculations for this paper have been supported by the
Supercomputing Center of Wuhan University.
We express our gratitude to our colleagues in the CERN
accelerator departments for the excellent performance of the LHC. We
thank the technical and administrative staff at the LHCb
institutes.
We acknowledge support from CERN and from the national agencies:
CAPES, CNPq, FAPERJ and FINEP (Brazil); 
MOST and NSFC (China); 
CNRS/IN2P3 (France); 
BMBF, DFG and MPG (Germany); 
INFN (Italy); 
NWO (Netherlands); 
MNiSW and NCN (Poland); 
MEN/IFA (Romania); 
MSHE (Russia); 
MICINN (Spain); 
SNSF and SER (Switzerland); 
NASU (Ukraine); 
STFC (United Kingdom); 
DOE NP and NSF (USA).
We acknowledge the computing resources that are provided by CERN, IN2P3
(France), KIT and DESY (Germany), INFN (Italy), SURF (Netherlands),
PIC (Spain), GridPP (United Kingdom), RRCKI and Yandex
LLC (Russia), CSCS (Switzerland), IFIN-HH (Romania), CBPF (Brazil),
PL-GRID (Poland) and OSC (USA).
We are indebted to the communities behind the multiple open-source
software packages on which we depend.
Individual groups or members have received support from
AvH Foundation (Germany);
EPLANET, Marie Sk\l{}odowska-Curie Actions and ERC (European Union);
A*MIDEX, ANR, Labex P2IO and OCEVU, and R\'{e}gion Auvergne-Rh\^{o}ne-Alpes (France);
Key Research Program of Frontier Sciences of CAS, CAS PIFI, CAS CCEPP, 
Fundamental Research Funds for the Central Universities, 
and Sci. \& Tech. Program of Guangzhou (China);
RFBR, RSF and Yandex LLC (Russia);
GVA, XuntaGal and GENCAT (Spain);
the Royal Society
and the Leverhulme Trust (United Kingdom).

\addcontentsline{toc}{section}{References}
\bibliographystyle{LHCb}
\bibliography{main,standard,LHCb-PAPER,LHCb-CONF,LHCb-DP,LHCb-TDR}
 
\section*{The LHCb experiment}

LHCb is one of the four big experiments located on the Larger Hadron Collider (LHC) at CERN.  
The LHCb detector is a single-arm forward spectrometer covering the pseudorapidity range $2 < \eta < 5$, designed for the study of particles containing \bquark\ or \cquark\ quarks. The detector includes a silicon-strip vertex detector surrounding the proton-proton interaction region, tracking stations on either side of a dipole magnet, ring-imaging Cherenkov (RICH) detectors, calorimeters and muon chambers.
The LHCb Collaboration consists of more than 1400 members from 18 countries in 5 continents, including both physicists and engineers.

\newpage
\centerline
{\large\bf LHCb Collaboration}
\begin
{flushleft}
\small
R.~Aaij$^{32}$,
C.~Abell{\'a}n~Beteta$^{50}$,
T.~Ackernley$^{60}$,
B.~Adeva$^{46}$,
M.~Adinolfi$^{54}$,
H.~Afsharnia$^{9}$,
C.A.~Aidala$^{85}$,
S.~Aiola$^{26}$,
Z.~Ajaltouni$^{9}$,
S.~Akar$^{65}$
% ...

{\footnotesize \it

% ...
$ ^{9}$Universit{\'e} Clermont Auvergne, CNRS/IN2P3, LPC, Clermont-Ferrand, France\\
% ...
$ ^{26}$INFN Sezione di Milano, Milano, Italy\\
% ...
$ ^{32}$Nikhef National Institute for Subatomic Physics, Amsterdam, Netherlands\\
% ...
$ ^{46}$Instituto Galego de F{\'\i}sica de Altas Enerx{\'\i}as (IGFAE), Universidade de Santiago de Compostela, Santiago de Compostela, Spain\\
% ...
$ ^{50}$Physik-Institut, Universit{\"a}t Z{\"u}rich, Z{\"u}rich, Switzerland\\
% ...
$ ^{54}$H.H. Wills Physics Laboratory, University of Bristol, Bristol, United Kingdom\\
% ...
$ ^{60}$Oliver Lodge Laboratory, University of Liverpool, Liverpool, United Kingdom\\
% ...
$ ^{65}$University of Cincinnati, Cincinnati, OH, United States\\
% ...
$ ^{68}$Syracuse University, Syracuse, NY, United States\\
% ...
$ ^{85}$University of Michigan, Ann Arbor, United States, associated to $^{68}$\\
\bigskip
% ...
\medskip
}
\end{flushleft}

\end{document}